\documentclass[conference]{IEEEtran}
\IEEEoverridecommandlockouts
\usepackage{cite}
\usepackage{amsmath,amssymb,amsfonts}
\usepackage{algorithm}
\usepackage{algorithmic}
\usepackage{graphicx}
\usepackage{textcomp}
\usepackage{xcolor}
\usepackage{subfig}
\usepackage{diagbox}
\usepackage{multirow}
\def\BibTeX{{\rm B\kern-.05em{\sc i\kern-.025em b}\kern-.08em
    T\kern-.1667em\lower.7ex\hbox{E}\kern-.125emX}}
\begin{document}

\title{Unsupervised clustering of disturbances in power systems via deep convolutional autoencoders\\

}

\author{\IEEEauthorblockN{ Md Maidul Islam, Md Omar Faruque}
\IEEEauthorblockA{\textit{Electrical and Computer Engineering} \\
\textit{FAMU-FSU College of Engineering},
Tallahassee, FL, USA \\
Email: mi19b@fsu.edu, mfaruque@eng.famu.fsu.edu}
\and
\IEEEauthorblockN{Joshua Butterfield, Gaurav Singh, Thomas A. Cooke}
\IEEEauthorblockA{\textit{Electric Power Research Institute} \\
Palo Alto, CA, USA \\
Email: {\{jbutterfield, gsingh, tcooke\}}@epri.com}
}

\maketitle

\begin{abstract}
Power quality (PQ) events are recorded by PQ meters whenever anomalous events are detected on the power grid. Using neural networks with machine learning can aid in accurately classifying the recorded waveforms and help power system engineers diagnose and rectify the root causes of problems. However, many of the waveforms captured during a disturbance in the power system need to be labeled for supervised learning, leaving a large number of data recordings for engineers to process manually or go unseen. This paper presents an autoencoder and K-means clustering-based unsupervised technique that can be used to cluster PQ events into categories like sag, interruption, transients, normal, and harmonic distortion to enable filtering of anomalous waveforms from recurring or normal waveforms. The method is demonstrated using three-phase, field-obtained voltage waveforms recorded in a distribution grid. First, a convolutional autoencoder compresses the input signals into a set of lower feature dimensions which, after further processing, is passed to the K-means algorithm to identify data clusters. Using a small, labeled dataset, numerical labels are then assigned to events based on a cosine similarity analysis. Finally, the study analyzes the clusters using the t-distributed stochastic neighbor embedding (t-SNE) visualization tool, demonstrating that the technique can help investigate a large number of captured events in a quick manner.  

\end{abstract}


\section{Introduction}
Power quality (PQ) events imply any variation in the voltage, current, or frequency supplied to a specific power system load. Modern-day electronic devices tend to be more sensitive to PQ disturbances than traditional loads, making PQ an important issue in maintaining a reliable power grid \cite{b1}. IEEE standard 1159 \cite{b2} defines seven broad categories of PQ events, each with several subcategories. The seven main categories are transients, short-duration variation, long-duration variation, voltage imbalance, waveform distortion, voltage fluctuation, and power frequency variation. When an anomalous event occurs, PQ meters may record voltage and current waveforms. To avoid the same event occurring again, it is of interest to utilities to analyze the recorded PQ event to assess its root causes so that corrective action may be taken to prevent more catastrophic damage or restore services faster. Such PQ data recordings are mostly unlabeled, meaning the cause of each recorded event is unknown, and each record includes event or non-event oscillography. As the waveform recordings are often performed periodically, and hundreds of anomalous events may occur within a network over a period of days, it can become overwhelming to identify the waveforms that require further analysis. 

With technological advances in power system monitoring devices, power system data with high resolution and sophisticated signal information are now available, enabling researchers to utilize artificial intelligence and machine learning techniques for analyzing PQ events. As such, the use of machine learning algorithms for the analysis of PQ waveforms is an area in which a significant body of work already exists \cite{OLIVEIRA2023108887}. Authors in \cite{b3} proposed a compressed sensing and deep convolution network-based PQ disturbance classification. An artificial neural network (ANN) and decision tree-based PQ event classification utilizing time and frequency domain features were presented in \cite{b4}. Unsupervised methods have also been utilized by researchers focusing on anomaly detection \cite{b5, b6} and event localization \cite{b8}. Most of these works are either based on synthetic labeled data or PMU data. Very little work has been performed on distribution grid analytics, mainly due to the lack of field-obtained data. Authors in \cite{b7} utilized Ward and K-means clustering method to cluster voltage sag utilizing $\mu$PMU recordings. A comparative review of signal processing and AI based methods of analyzing PQ events in smartgrid is presented in \cite{b9}. The Electric Power Research Institute (EPRI) is in the process of constructing a PQ event waveform library \cite{pqmonlib}, in which a large number of PQ event waveforms are to be consolidated for use in future research. As most of the waveforms in this library are unlabeled, it is desired to create a data-driven clustering method capable of grouping different event data into clusters and utilizing a handful of labeled data, indicating the event type of each cluster. 
The main contributions of this paper are:
\begin{itemize}
    \item Development of an autoencoder and K-means-based unsupervised algorithm to cluster unlabeled field-obtained waveform data.
    \item Utilization of cosine similarity between cluster centers and a small number of labeled data to identify different PQ event labels for the clusters.
    \item An analysis of t-SNE visualization output to interpret the different clusters.
\end{itemize}
Utilizing these methods presents novelty first by addressing the need to isolate ``interesting" events from recurring events. Over time, this approach can be trained by PQ engineers to automatically prioritize data for further analysis based on the cluster in which the event is classified and whether the event has high similarity to other recorded events. Secondly, this paper helps to address the lack of research performed on field-collected data, as the majority of available sources utilize modeled or synthetic waveform data.

The remainder of the paper is organized as follows: section II presents the autoencoder and K-means-based clustering methodology, section III discusses clustering performances and cosine similarity-based event label identification, and finally, section IV concludes the paper.

\section{Unsupervised clustering method}

\subsection{PQ event data preprocessing}
\begin{figure}[ht!]
\centering
\includegraphics[height=6.5cm,width=\linewidth]{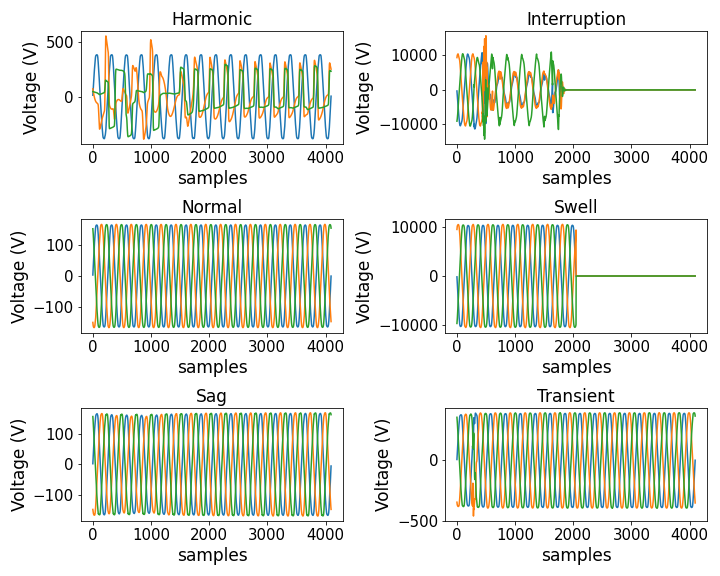}
\caption{Voltage waveform recording from a distribution utility.}
\label{fig:pqevent}
\end{figure}

The EPRI waveform database consists of voltage and current waveform recordings obtained from different physical locations within a distribution system. These waveforms have been recorded periodically by PQ meters. The data library contains normal and PQ event data without any labels or identifying information attached. In this work, 5112 unlabeled voltage waveform recordings have been used. Of these waveforms, 250 waveforms of different categories have been labeled based on manual observation. Fig. \ref{fig:pqevent} shows some of the recorded events from the manually labeled data. 

\begin{figure*}[ht!]
\centering
\includegraphics[height=5cm,width=0.95\textwidth]{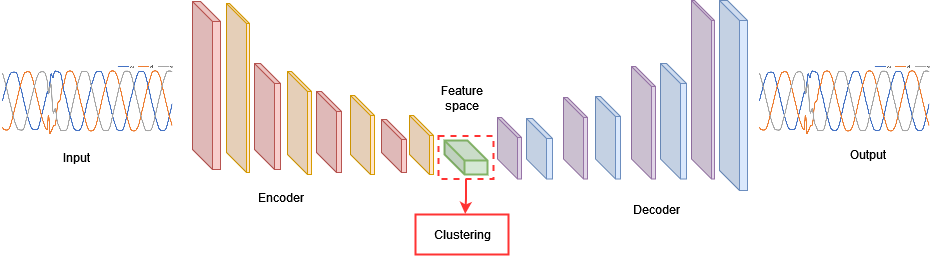}
\caption{Convolutional autoencoder structure}
\label{fig:ae_struct}
\end{figure*}
The recorded waveforms are 16 cycles long, with 256 samples in each cycle. Thus, for three-phase voltage samples, total 4096$\times$3 points are recorded in a waveform capture. Some of the data were 8 cycles long, i.e, the swell event example in Fig. \ref{fig:pqevent}, and zero padding was performed to ensure that all waveform recordings had the same length. Per-unit scaling was applied to the data, and any data less than 8 cycles in length were removed. It should be noted that only voltage waveforms have been utilized as they better depict some of the PQ events like swell and sag. It is well known that dealing with high-dimensional data is a significant challenge for the application of clustering. Thus, applying dimensionality reduction and performing clustering in feature space instead of data space improves the clustering performance \cite{BOUVEYRON2007502}. A deep convolutional autoencoder has been applied to these waveforms to learn the salient representation of data in a lower-dimensional feature space. 

\subsection{Feature space formation via deep convolutional autoencoder}

Autoencoders (AE) are a form of unsupervised representation learning in which the neural network learns to reconstruct an input signal at the network output \cite{goodfellow2016deep}. There are two stages in an AE; in the first stage, the encoder learns to represent the input data in a latent space, and the decoder stage aims at reconstructing the input data from the latent space so that the reconstruction error is minimum. Fig. \ref{fig:ae_struct} shows the structure of the convolutional AE. In this work, convolutional AE has been applied to the input waveforms, and the encoder's output has been utilized as the feature space for the clustering algorithm. This method is known as under complete AE. Let, $f_W(.)$ represent the encoder network and $g_U(.)$ the corresponding decoder. The AE aims at minimizing the reconstruction mean squared error (MSE) between input, $x$, and output:

\begin{equation}
\min\limits_{W, U}\frac{1}{n}\sum_{i=1}^{n} \left\|g_U(f_W(x_i))-x_i\right\|_2^2
\end{equation}

Where, for convolutional AE, the encoder and decoder network is described by:

\begin{align}
f_W(x) =& \sigma (x * W + b) \equiv h \\
g_U(h) =&  \sigma(h * U + b')
\end{align}

Here, $\sigma$ is a nonlinear activation function, h is the feature space, `*' is the convolution operator, W, U are the weights, and b, b' are biases corresponding to the encoder and decoder network. The AE learns the salient representation of the waveforms in feature space, $h$, by minimizing the reconstruction error. 

\subsection{K-means clustering}
K-means is a well-known clustering algorithm that uses distance-based metrics to identify similar data samples \cite{9072123}. Let the dataset, $x= {x^{(1)}, ..., x^{(m)}}$, be represented by feature set $h= {h^{(1)}, ..., h^{(m)}}$ in a d-dimensional Euclidean space $\mathbb{R}^d$, in which the goal is to group the data into a few cohesive ``clusters". Let $a = {a_1, a_2,....,a_c}$ be the c cluster centers, and $z = [z_{ik}]_{n \times c}$ where $z_{ik} \in {0, 1}$ indicates if $x_i \in k_{th}$ cluster. The k-means algorithm thus tries to minimize the objective function $J(z,A) = \sum_{i=1}^{n}\sum_{k=1}^{c}\left\|x_i-a_k\right\|^2$. The algorithm is iterated until boundary conditions are met. The cluster centers and memberships are updated respectively as:
\begin{align}
a_k = & \frac{\sum_{i=1}^{n}z_{ik}x_{i}}{\sum_{i=1}^{n}z_{ik}} \\
z_{ik}=& \begin{cases}
      1, & \text{if}\ \left\|x_i-a_k\right\|^2 = \min\limits_{1 \leq k\leq c} \left\|x_i-a_k\right\|^2\\
      0, & \text{otherwise}
    \end{cases}  
\end{align}
Here, $\left\|x_i-a_k\right\|^2$ is the Euclidean distance between data point $x_i$ and cluster center $a_i$. It should be noted that although K-means is an unsupervised algorithm, it requires a priori information of the number of clusters. The number of clusters can be selected based on expert knowledge of the dataset or from some well-known cluster selection methods, such as the elbow method or silhouette score.

\subsection{Parameter selection and tuning}
\begin{figure}[ht]
\begin{center}
    \subfloat[Evolution of training and validation loss with epoch]{\label{fig:learning_curve}\includegraphics[width=0.95\linewidth]{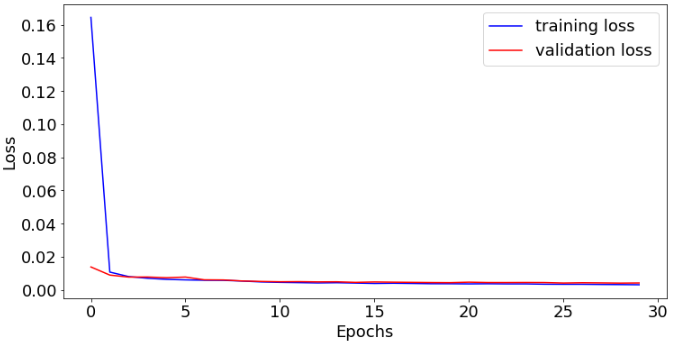}} \\
    \subfloat[Reconstruction error]{\label{fig:recon_error}\includegraphics[width=0.95\linewidth]{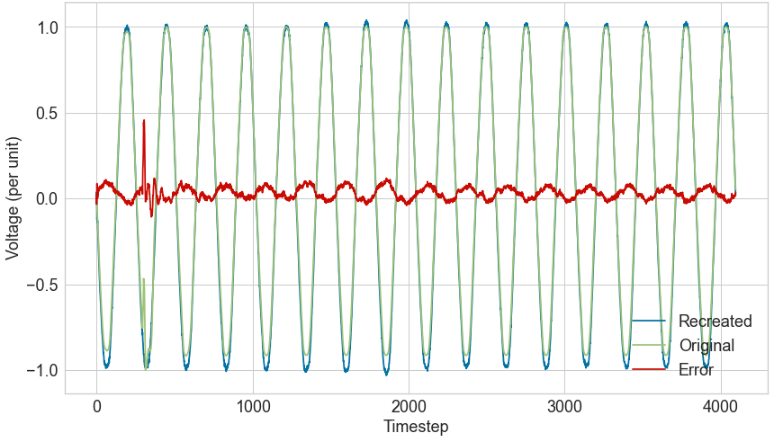}}
    \caption{Auto encoder learning}
    \end{center}
\end{figure}

In the next part of this work, the three-phase voltage data after preprocessing was fed to the AE for feature extraction. The data samples were split in 70:15:15 ratio into training, validation, and test data. The AE is composed of five convolutional layers followed by five pooling layers and finally, the features are flattened to embed in 60 features. A dense layer is followed by three convolutional layers in the decoder part.  Fig. \ref{fig:learning_curve} shows the training and validation loss with each epoch, and fig. \ref{fig:recon_error} shows the reconstruction error for sample test data. 

\begin{figure}[ht!]
\centering
\includegraphics[width=\linewidth]{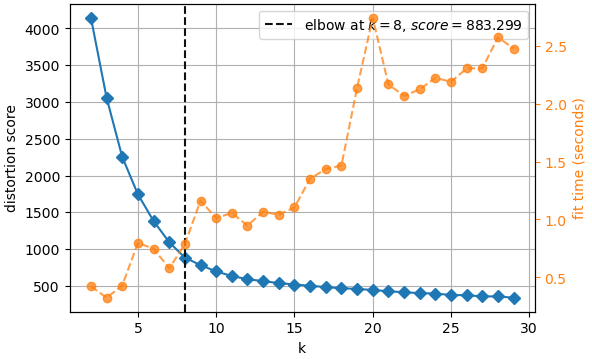}
\caption{Cluster number selection by elbow method}
\label{fig:elbow}
\end{figure}

The overall clustering performance depends on both the reconstruction error in the AE and the clustering error in the K-means algorithm. While it is easier to embed the data into a higher dimension, K-means algorithm accuracy suffers from the high dimensionality of the feature space data. A grid search method was applied to find the best results, and a feature set of 60 yielded the overall best clustering performance. After encoding the data, principal component analysis (PCA) was applied to keep $95\%$ variance in the feature set. This further reduces the feature set to 17 features and improves the performance of clustering. Although K-means is an unsupervised algorithm, it requires the cluster number as an input variable. In this paper, within-cluster-sum of squared errors (WSS), also known as `distortion score', has been used as a metric to find the ideal number of clusters. WSS depicts the variability of samples within each cluster and is defined as:
\begin{equation}
    WSS= \sum_{k=1}^{c}\sum_{x_i \in k}\left\|x_i-a_k\right\|^2
\end{equation}
Based on the WSS score for different number of clusters (k), the suitable k can be detected by the elbow method. Fig. \ref{fig:elbow} shows that WSS changes rapidly for the low number of clusters and slows down for a  higher k forming the elbow at k=8. However, the cluster number is a tunable parameter, and if the elbow is not well-defined, different cluster sizes near the elbow point should be explored. The K-means algorithm randomly initializes k cluster centroids and updates cluster members and cluster centroids based on the distance. The output of the algorithm is cluster numeric labels associated with each data input.   

\section{Results and discussion}

\begin{figure}[ht!]
\centering
\includegraphics[height=7.5cm,width=\linewidth]{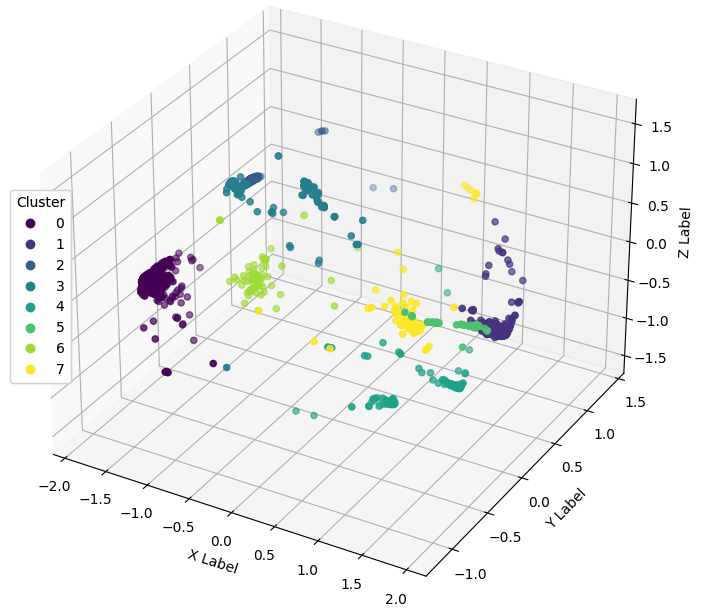}
\caption{Scatterplot in 3d with K-means}
\label{fig:scatter_3d}
\end{figure}

Fig. \ref{fig:scatter_3d} presents a three-dimensional scatter plot based on the output from the K-means algorithm. PCA was applied to the feature set to reduce it to three major components to be able to visualize in a 3d plot. The data is segmented into eight clusters represented by numerical labels. 

\begin{figure}[ht!]
\centering
\includegraphics[height=5.5cm, width=\linewidth]{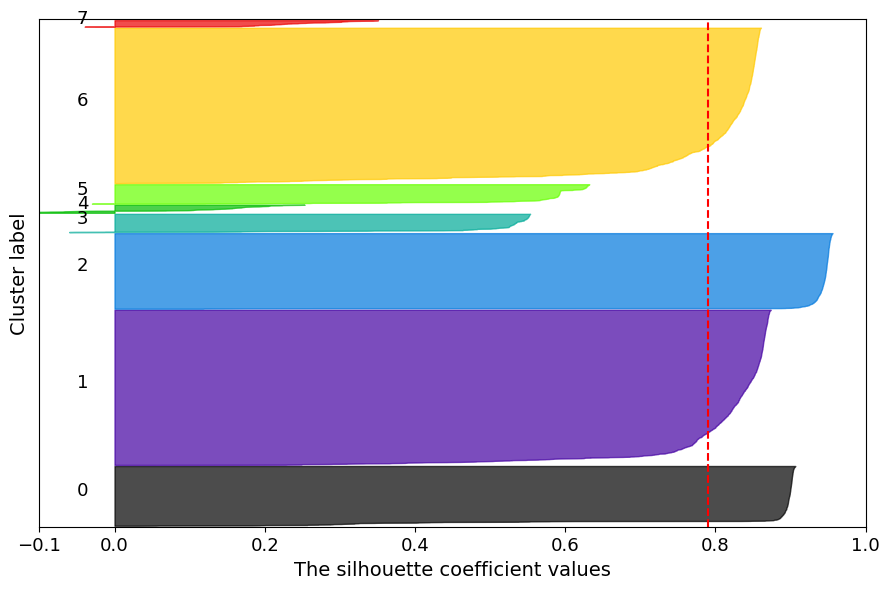}
\caption{Silhouette analysis for K-means clustering}
\label{fig:silhouette}
\end{figure}

The cluster sizes are imbalanced, which depicts that the frequency of different power events in a distribution system is also imbalanced. Additionally, some different events share some common features; thus, it is highly unlikely to get completely separated clusters. Unlike supervised learning, there is no ideal metric for evaluating cluster performance in an unsupervised clustering method. Silhouette score \cite{ROUSSEEUW198753} is often used as a metric to measure the integrity and quality of the clusters. Silhouette coefficient s(i) for a sample i is defined as:

\begin{equation}
    s(i)=\frac{b(i)-a(i)}{max(b(i), a(i))}
\end{equation}
Here, a(i) is the average distance between each point within a cluster, and b(i) is average inter-cluster distance or the average distance between all clusters, s(i)$\in$[-1,1], where a value close to 1 means higher similarity within cluster and good separation from neighbouring clusters. Fig. \ref{fig:silhouette} presents silhouette analysis for the clustering performance with cluster sizes represented by thickness for each cluster label. The average silhouette score is 0.789, which represents cluster integrity and quality. 

In addition to creating the clusters, it is highly desirable to identify the events corresponding to each cluster. Some of the labeled data in the repository have been utilized to identify the events. Manual observation of the cluster samples can be used to identify the different PQ events if there is no labeled data. Cosine similarity has been used to identify similarity among the cluster centroids and labeled data. Cosine similarity, $c_{sim}$ is defined as:
\begin{equation}
    c_{sim}= \frac{a_k \cdot x_i}{\left\|a_k\right\|\left\|x_i\right\|}
\end{equation}

Algorithm \ref{algo} shows the process of assigning labels to clusters. Cosine similarity with cluster centroids from the K-means algorithm is calculated for each labeled sample. Then mean cosine similarity, $c_{sim}$ for each event is calculated, and the cluster centroid with maximum $c_{sim}$ identifies that event label. 

\begin{algorithm}[]
\caption{Event assign to clusters via cosine similarity}
\label{algo}
\begin{algorithmic}[1]
    \FOR{Cluster centers, $a_k\in\{a_1, a_2,...,a_c\}$}
    \STATE{similarity, s=[]
      \FOR{$x\in\{x_1,...,x_{ik}\}$}
     \STATE{Encode data, $h_i=f_w(x_i)$ }\\
      \STATE{Apply PCA }\\
      \STATE{Calculate $c_{sim}$ = cosine similarity($a_k,h_i$)}\\
      \STATE{Append $c_{sim}$ to s}
      \ENDFOR
    }
    \STATE{Compute mean similarity, $s_{mean}$}
    \STATE{Assign event label to $a_k$, for $max(s_{mean})$}
    \ENDFOR
\end{algorithmic}
\end{algorithm}

\begin{figure}[ht]
\centering
\includegraphics[height=6cm,width=\linewidth]{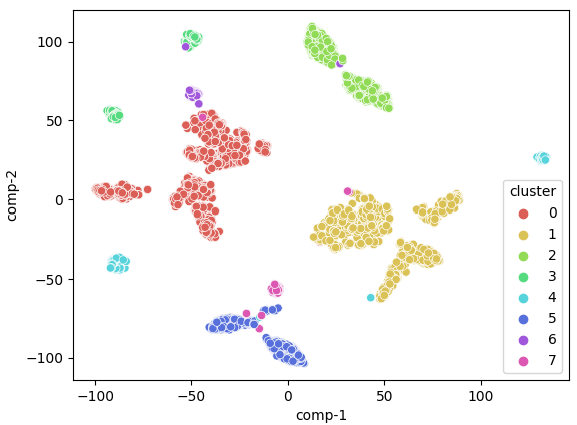}
\caption{t-SNE visualization of clusters}
\label{fig:tsne}
\end{figure}

In the labeled data, there were five types of event samples. Table \ref{tab:label} shows the average cosine similarity of each cluster center with the labeled data and probable events tagged with numerical labels associated with the clusters. The label assignment is affected by the accuracy and amount of labeled data. Insufficient data for an event can create ambiguous results. Also, there can be a high similarity between several events. In those cases, manual observation of some of the samples in the cluster can be performed to identify the correct associated event. From this analysis, it is evident that there could be some variation within each cluster. For example, there can be events that have both transient and sag characteristics, or there could be events of voltage swell with harmonics.  
\begin{table}[ht!]
\centering
\caption{Assign label based on similarity with labeled data}
\label{tab:label}
\resizebox{\linewidth}{!}{
\begin{tabular}{|l||*{6}{c|}}
\hline
\multirow{2}{*}{\diagbox[width=6em]{Cluster}{Event}}& \multicolumn{6}{c|}{Cosine similarity}  \\ \hline 
& Transients & Harmonics & Normal & Sag & Interruption & Event  \\\hline\hline
Cluster 0 & 0.40327 & 0.2950542 & 0.3583036 & 0.5466619 & 0.35973388 & Sag \\ \hline
Cluster 1 & 0.63310885 & 0.6049854 & 0.4826975 & 0.5003011 & 0.5400346 & Transients \\ \hline
Cluster 2 & 0.59648067 & 0.57894146 & 0.67399657 & 0.473256 & 0.602269 & Normal \\ \hline
Cluster 3 & 0.53878766 & 0.72971654 & 0.5361408 & 0.5481992 & 0.82220125 & Interruption \\ \hline
Cluster 4 & 0.5984593 & 0.5690439 & 0.5972265 & 0.4701643 & 0.6424283 & Cluster 4 \\ \hline
Cluster 5 & 0.4082755 & 0.38652408 & 0.45341122 & 0.38993802 & 0.39789283 & Normal \\ \hline
Cluster 6 & 0.49006578 & 0.49370754 & 0.47652954 & 0.5237515 & 0.6401432 & Cluster 6 \\ \hline
Cluster 7 & 0.5691603 & 0.7461975 & 0.59154147 & 0.47507018 & 0.7865461 & Harmonics \\ \hline
\end{tabular}
}
\end{table}
With the cluster labels updated, the t-distributed stochastic neighbor embedding (t-SNE) algorithm \cite{van2008visualizing} was applied to the feature space. t-SNE is a dimensionality reduction technique for the visualization of high-dimensional data. Fig. \ref{fig:tsne} presents t-SNE visualization, which indeed shows some local variations. 

It provides further insights into the clusters. However, as t-SNE does not preserve distance or density, the cluster size, shape, or distances do not represent sample characteristics. Variations within clusters like single phase affected vs. two phases affected (cluster 3), sag with transient, sustained sag vs. temporary sag can be observed by investigating the subclusters. In addition, some incorrect cluster assignments can be observed, especially in cases where data with smaller lengths have been wrongly clustered as interruptions. A possible way to improve the performance would be to maintain the same length and format of data capturing. Ultimately, this method can be used to filter the massive amount of recorded PQ waveforms so that uninteresting data are removed. By clustering different events and using cosine similarity, it is envisioned that the operator could manually identify a small number of waveforms as ``interesting" or ``uninteresting," and the algorithm could respond by alerting the operator to the appearance of more ``interesting" events on the system.

\section{conclusion}
Unsupervised clustering can be a useful tool to visualize and identify data of interest in the context of power system data analysis. In this paper, an autoencoder-based unsupervised waveform clustering method has been explored to group a large number of unlabeled data into different clusters. The selection of cluster number and performance analysis has been discussed. Using some available labeled data, a cosine similarity-based cluster label assigning method has been presented to identify different PQ events. Finally, some visualization methods and cluster analysis have been discussed. The proposed framework would be instrumental in analyzing unlabeled waveform captures and aid power system engineers in taking appropriate actions for maintaining power quality.

\vspace{12pt}
\color{red}


\begin{thebibliography}{00}
\bibitem{b1}Chawda, G., Shaik, A., Shaik, M., Padmanaban, S., Holm-Nielsen, J., Mahela, O. \& Kaliannan, P. Comprehensive Review on Detection and Classification of Power Quality Disturbances in Utility Grid With Renewable Energy Penetration. {\em IEEE Access}. \textbf{8} pp. 146807-146830 (2020)
\bibitem{b2} A IEEE Recommended Practice for Monitoring Electric Power Quality. {\em IEEE Std 1159-2019 (Revision Of IEEE Std 1159-2009)}. pp. 1-98 (2019)
\bibitem{OLIVEIRA2023108887}Oliveira, R. \& Bollen, M. Deep learning for power quality. {\em Electric Power Systems Research}. \textbf{214} pp. 108887 (2023)
\bibitem{b3} Wang, J., Xu, Z. \& Che, Y. Power Quality Disturbance Classification Based on Compressed Sensing and Deep Convolution Neural Networks. {\em IEEE Access}. \textbf{7} pp. 78336-78346 (2019)
\bibitem{b4} Aligholian, A., Farajollahi, M. \& Mohsenian-Rad, H. Unsupervised Learning for Online Abnormality Detection in Smart Meter Data. {\em 2019 IEEE Power \& Energy Society General Meeting (PESGM)}. pp. 1-5 (2019)
\bibitem{b5} Pandey, S., Srivastava, A. \& Amidan, B. A Real Time Event Detection, Classification and Localization Using Synchrophasor Data. {\em IEEE Transactions On Power Systems}. \textbf{35}, 4421-4431 (2020)
\bibitem{b6} Aligholian, A., Shahsavari, A., Stewart, E., Cortez, E. \& Mohsenian-Rad, H. Unsupervised event detection, clustering, and use case exposition in micro-pmu measurements. {\em IEEE Transactions On Smart Grid}. \textbf{12}, 3624-3636 (2021)


\bibitem{b8} Li, H., Weng, Y., Farantatos, E. \& Patel, M. An Unsupervised Learning Framework for Event Detection, Type Identification and Localization Using PMUs Without Any Historical Labels. {\em 2019 IEEE Power \& Energy Society General Meeting (PESGM)}. pp. 1-5 (2019)

\bibitem{b7} Swenson, T., Vrettos, E., Müller, J. \& Gehbauer, C. Open µPMU Event Dataset: Detection and Characterization at LBNL Campus. {\em 2019 IEEE Power \& Energy Society General Meeting (PESGM)}. pp. 1-5 (2019)

\bibitem{b9} Beniwal, R., Saini, M., Nayyar, A., Qureshi, B. \& Aggarwal, A. A Critical Analysis of Methodologies for Detection and Classification of Power Quality Events in Smart Grid. {\em IEEE Access}. \textbf{9} pp. 83507-83534 (2021)

\bibitem{pqmonlib}DOE/EPRI National Database Repository of Power System Events (https://pqmon.epri.com/)

\bibitem{BOUVEYRON2007502}Ali, M., Alqahtani, A., Jones, M. \& Xie, X. Clustering and Classification for Time Series Data in Visual Analytics: A Survey. {\em IEEE Access}. \textbf{7} pp. 181314-181338 (2019)

\bibitem{goodfellow2016deep}Goodfellow, I., Bengio, Y. \& Courville, A. Deep learning. (MIT press,2016)

\bibitem{9072123}Sinaga, K. \& Yang, M. Unsupervised K-Means Clustering Algorithm. {\em IEEE Access}. \textbf{8} pp. 80716-80727 (2020)

\bibitem{ROUSSEEUW198753}Rousseeuw, P. Silhouettes: A graphical aid to the interpretation and validation of cluster analysis. {\em Journal Of Computational And Applied Mathematics}. \textbf{20} pp. 53-65 (1987)

\bibitem{van2008visualizing}Maaten, L. \& Hinton, G. Visualizing data using t-SNE{ \em Journal Of Machine Learning Research}. \textbf{9} (2008)

\end{thebibliography}
\end{document}